\documentclass[letter,a4paper,twocolumn]{jpsj2}
\title{
	Self-Diffusion of a Polymer Chain in a Melt
}
\author{
	Katsumi {\sc Hagita}
	and
	Hiroshi {\sc Takano}
}
\def\runauthor{
	Katsumi {\sc Hagita} and
	Hiroshi {\sc Takano}
}
\inst{
	Department of Physics,
	Faculty of Science and Technology,\\
	Keio University,
	Yokohama 223-8522
}
\recdate{\hspace*{5cm}}
\abst{%
Self-diffusion of a polymer chain in a melt
is studied by Monte Carlo simulations of the bond fluctuation model,
where only the excluded volume interaction is taken into account.
Polymer chains, each of which consists of $N$ segments,
are located on an $L \times L \times L$
simple cubic lattice under periodic boundary conditions,
where each segment occupies $2 \times 2 \times 2$ unit cells.
The results for $N=32, 48, 64, 96, 128, 192, 256, 384$ and $512$
at the volume fraction $\phi \simeq 0.5$
are reported,
where $L = 128$ for $N \leq 256$ and $L=192$ for $N \geq 384$.
The $N$-dependence of the self-diffusion constant $D$
is examined.
Here, $D$ is estimated from the
mean square displacements of the center 
of mass of a single polymer chain
at the times larger than the longest relaxation time.
From the data for $N = 256$, $384$ and $512$,
the apparent exponent $x_{\rm d}$,
which describes the apparent power law dependence of $D$
on $N$ as $D \propto N^{- x_{\rm d}}$,
is estimated as $x_{\rm d} \simeq 2.4$.
The ratio
$D \tau / \langle R_{\rm e}^{2} \rangle$
seems to be a constant
for $N = 192, 256, 384$ and $512$,
where
$\tau$
and
$\langle R_{\rm e}^{2} \rangle$
denote
the longest relaxation time
and
the mean square end-to-end distance,
respectively.
}
\kword{
	polymer chain,
	melt,
	self-diffusion,
	diffusion constant,
	Monte Carlo simulations,
	lattice model,
	bond fluctuation model,
	reptation
}
\begin{document}
\sloppy
\maketitle
\par
Recently, many works have been done
in order to study
the slow dynamics of a single polymer chain in a melt
through simulations.%
\cite{%
HagitaTakano:melt,
Binder,%
PaulBinder,%
PaulBinder2,%
KremerGrest,
KremerGrest2,%
KreerBaschnagelMullerBinder,%
DunwegGrestKremer,%
Kaznessis2%
}
These studies are motivated by
the reptation theory
of the dynamics of concentrated polymer systems.%
\cite{%
DoiEdwards,%
deGennes%
}
In the reptation theory,
the dynamics of concentrated polymer systems
is described in terms of
the motion of a polymer chain
trapped in a tube,
which represents the entanglement effects of the surrounding polymers.
The self-diffusion constant $D$ of the center of mass
and
the longest relaxation time $\tau$
of a polymer chain of $N$ segments
are predicted to behave as
$D \propto N^{-2} $
and 
$\tau \propto N^{3} $,
respectively.
On the other hand,
the experimental results on the behaviors of
$D$ and $\tau$
have been summarized as
$D \propto N^{-2}$ and $\tau \propto N^{3.4}$.%
\cite{DoiEdwards,deGennes}
Thus, the exponent for $\tau$ observed in the experiments
is different from that predicted by the reptation theory.
This discrepancy between the experiments and the reptation theory
is considered to be attributed to
the contour length fluctuation.%
\cite{%
DoiEdwards,%
Doi:clf,%
OConnorBall,%
MilnerMcLeish%
}
The contour length fluctuation makes
the relaxation time $\tau$
shorter than that predicted by the original reptation theory.
As $N/N_{\rm e}$ becomes very large,
the effect of the contour length fluctuation becomes negligible
and
$\tau$ approaches the predicted power law behavior $N^3$ from below,
where $N_{\rm e}$ denotes the number of segments
between entanglement points.
Because of this crossover behavior,
the apparent exponent for $\tau$
increases beyond three and reaches $3.4$
for intermediately large values of $N/N_{\rm e}$.
This behavior of the apparent exponent for $\tau$
has been confirmed
by the present authors
through Monte Carlo simulations.%
\cite{HagitaTakano:melt}
\par
In contrast to the relaxation time $\tau$,
the exponent for the self-diffusion constant 
$D$
observed in the experiments
has been believed to agree with 
that predicted by the reptation theory.
Therefore, 
the contour length fluctuation has been considered
to have little effect on
the $N$-dependence of the self-diffusion constant
$D$.
In the recent experiments by Tao, Lodge and Meerwall,%
\cite{Lodge,TaoLodgeMeerwall}
however,
it is reported that
the self-diffusion constant $D$
in hydrogenated polybutadiene (hPB) alkane solutions
and hPB melt behaves as
$D \propto N^{-2.4 \pm 0.1}$.
Theoretically,
Frischknecht and Milner%
\cite{FrischknechtMilner}
reported that
the apparent power law dependence of $D$ on $N$
is modified by the contour length fluctuation
as $D \propto N^{-2.25}$
at $N/N_{\rm e} \simeq 80$,
while it approaches
the behavior $D \propto N^{-2}$
predicted by the reptation theory
as $N/N_{\rm e}$ becomes larger.
Thus, the behavior of the self-diffusion constant $D$
is still in dispute.
Therefore, it is desirable
to elucidate the $N$-dependence of $D$
through simulations.
Even in the recent simulations of a polymer melt,%
\cite{%
KreerBaschnagelMullerBinder,%
DunwegGrestKremer
}
however,
the $N$-dependence of $D$ has not been clarified.
D\"{u}nweg, Grest and Kremer%
\cite{DunwegGrestKremer}
performed molecular dynamics simulations
for $N=5, 10, 25, 40, 50, 100, 200$ and $350$
at the reduced density $\rho = 0.85$
and the reduced temperature $k_{\rm B}T = 1$.%
\cite{KremerGrest,KremerGrest2}
They estimated $D$ only for $N \le 200$.
From their values of $D$ for $N=100$ and $200$,
the apparent exponent $x_{\rm d}$
in the  apparent power law $D \propto N^{-x_{\rm d}}$
can be estimated as
$x_{\rm d} \simeq 1.93$.
However, 
the values $N=100$ and $200$ are not in the range where
the apparent exponent for the relaxation time $\tau$ exceeds three.%
\cite{DunwegGrestKremer}
Therefore, their results cannot resolve the dispute
as to the apparent power law dependence of $D$ on $N$.
Moreover, the sizes of the systems used in their simulations
do not seem to be sufficiently large.
The ratio of the linear dimension of 
the simulation box to the root mean square end-to-end
distance of a polymer chain is
$1.76$ and $1.54$ for $N=100$ and $200$, respectively.
Kreer, Bashnagel, M\"uller and Binder%
\cite{%
KreerBaschnagelMullerBinder%
}
performed Monte Carlo simulations of the bond fluctuation model%
\cite{CarmesinKremer}
for $N = 16, 32, 64, 128$ and $512$
at the volume fraction $\phi = 0.5$.%
\cite{PaulBinder}
From their values of $D$ for  $N=128$ and $512$,
the apparent exponent $x_{\rm d}$
is estimated to be a little smaller than two.
Although the values $N=128$ and $512$
are in the range where
the apparent exponent for the relaxation time exceeds three,%
\cite{%
HagitaTakano:melt,%
KreerBaschnagelMullerBinder%
}
they are only two values of $N$ and 
it is not clear
how the apparent exponent $x_{\rm d}$ 
changes as $N$ becomes large.
Furthermore,
the problem of the system size also exists in their simulations.
The ratio of the linear dimension of the system
to the root mean square end-to-end distance
is $1.75$ for $N=512$.
\par
The purpose of the present letter is
to clarify the $N$-dependence of $D$
in the range of values of $N$
where the apparent exponent for the relaxation time exceeds three.
As in our previous letter,%
\cite{HagitaTakano:melt}
we perform Monte Carlo simulations of 
the same model as used in Ref.\ %
\citen{KreerBaschnagelMullerBinder}%
.
More values of $N$ than in  Ref.\ %
\citen{KreerBaschnagelMullerBinder}
are studied.
For each value of $N$,
the linear dimension of the simulation box is at least
$2.5$ times larger than
the root mean square end-to-end distance.
The self-diffusion constant $D$
is estimated
from the mean square displacements of
the center of mass of a polymer chain
at the times larger than the longest relaxation time.%
\cite{HagitaTakano:melt}
\par
In the present study,
the bond fluctuation model%
\cite{CarmesinKremer}
is used for Monte Carlo simulations of a polymer melt.
A system of $M$ polymer chains,
each of which consists of $N$ segments, is considered
on a simple cubic lattice
of which the lattice constant is $a$.
Each side of the lattice has $L$ lattice points
and the periodic boundary conditions are applied.
The polymer chains are represented by $M$ series of $N$ lattice points
$\{ (\mib{r}_{1,m}, \cdots, \mib{r}_{N,m}) ;\ m = 1, \cdots, M \}$,
where
$\mib{r}_{i,m}$ denotes
the center of the $i$th segment of the $m$th polymer chain.
Each segment occupies a cube of $2^3$ unit cells
and no overlap of segments is allowed.
The volume fraction
is given by
$\phi = 2^{3} M N / L^{3}$.
The lengths of the bond vectors
are restricted to the five prescribed lengths
$\{ 2a, \sqrt{5}a, \sqrt{6}a, 3a, \sqrt{10}a \}$.
The center of each segment stochastically
moves to one of its nearest neighbor lattice points
with the transition probability per unit time $\gamma / 6$
if the new position is allowed by
the conditions for the excluded volume and the bond vectors.
The dynamics is simulated by
a standard Monte Carlo method,
which uses discrete time steps
and updates $MN$ segments in random sequence.
The time scale $\gamma^{-1}$
corresponds to
one update per segment.
\cite{KosekiHiraoTakano}
We choose $a=1$ and $\gamma = 1$.
\par
In the present study, 
the cases of the volume fraction  $\phi \simeq 0.5$
are examined
for $N=32, 48, 64, 96, 128, 192, 256, 384$ and $512$.
We choose $L=128$ for $N \leq 256$ and $L=192$ for $N \geq 384$. 
For these values of $L$,
it is confirmed that no polymer chain interacts with 
its images generated by
the periodic boundary conditions in the present simulations.
The vectorized code for the bond fluctuation model%
\cite{WittmannKremer}
is used on vector processors for $N \geq 192$.
The equilibrium average is
calculated as the average of
the time averages obtained from
Monte Carlo simulations of four statistically independent systems.
In each Monte Carlo simulation,
the time average is calculated from the sequence of
$N_{\rm I}$ states
taken at intervals of $t_{\rm I}$ after
the initial period of $T_{\rm i}$,
which is discarded for the equilibration.
The values of the parameters used in the present simulations are shown
in Table \ref{t-values}.
\begin{table*}[tb]
\caption{
The parameters of the simulations.
The mean square end-to-end distance
$\langle R_{\rm e}^2 \rangle$,
the inverse of the longest relaxation time $1/\tau$
and
the self-diffusion constant $D$
of a polymer chain
are also shown.
The values of 
$\langle R_{\rm e}^2 \rangle$
and
$1/\tau$
for $N \le 256$
are taken from Ref.\ 
\citen{HagitaTakano:melt}.
The symbols k and M denote $\times 10^3$ and $\times 10^6$,
respectively.
}
\label{t-values}
\begin{tabular*}{\textwidth}{
@{\hspace{\tabcolsep}}
*{3}{r@{\extracolsep{\fill}}}
l@{\extracolsep{\fill}}
*{8}{r@{\extracolsep{\fill}}}
r
@{\hspace{\tabcolsep}}
r
@{\hspace{\tabcolsep}}
}
\hline
\multicolumn{1}{@{\hspace{\tabcolsep}}c@{\extracolsep{\fill}}}{$N$}
& \multicolumn{1}{@{\extracolsep{\fill}}c@{\extracolsep{\fill}}}{$M$}
& \multicolumn{1}{@{\extracolsep{\fill}}c@{\extracolsep{\fill}}}{$L$}
& \multicolumn{1}{@{\extracolsep{\fill}}c@{\extracolsep{\fill}}}{$\phi$}
& \multicolumn{1}{@{\extracolsep{\fill}}c@{\extracolsep{\fill}}}{$T_{\rm i}$}
& \multicolumn{1}{@{\extracolsep{\fill}}c@{\extracolsep{\fill}}}{$t_{\rm I}$}
& \multicolumn{1}{@{\extracolsep{\fill}}c@{\extracolsep{\fill}}}{$N_{\rm I}$}
& \multicolumn{1}{@{\extracolsep{\fill}}c@{\extracolsep{\fill}}}{$\langle R_{\rm e}^2 \rangle$}
& \multicolumn{1}{@{\extracolsep{\fill}}c@{\hspace{\tabcolsep}}}{$1/\tau$}
& \multicolumn{1}{@{\extracolsep{\fill}}c@{\hspace{\tabcolsep}}}{$D$}
\\
\hline
  $32$
& $4096$
& $128$
& $0.5$
& $0.5$M
& $0.5$k
& $16$k
& $292$
& $1.33 \times 10^{-5}$
& $1.27 \times 10^{-4}$
\\
  $48$
& $2720$
& $128$
& $0.498$
& $0.5$M
& $0.5$k
& $16$k
& $451$
& $5.03 \times 10^{-6}$
& $7.11 \times 10^{-5}$
\\
  $64$
& $2048$
& $128$
& $0.5$
& $2$M
& $1$k
& $32$k
& $614$
& $2.43 \times 10^{-6}$
& $4.44 \times 10^{-5}$
\\
$96$
& $1360$
& $128$
& $0.498$
& $2$M
& $5$k
& $1.6$k
& $939$
& $8.63 \times 10^{-7}$
& $2.33 \times 10^{-5}$
\\
$128$
& $1024$
& $128$
& $0.5$
& $5$M
& $5$k
& $4$k
& $1261$
& $3.98 \times 10^{-7}$
& $1.30 \times 10^{-5}$
\\
$192$
& $680$
& $128$
& $0.498$
& $20$M
& $20$k
& $4$k
& $1919$
& $1.20 \times 10^{-7}$
& $5.65 \times 10^{-6}$
\\
$256$
& $512$
& $128$
& $0.5$
& $40$M
& $20$k
& $8$k
& $2570$
& $4.65 \times 10^{-8}$
& $2.78 \times 10^{-6}$
\\ 
$384$
& $1152$
& $192$
& $0.5$
& $100$M
& $0.4$M
& $1$k
& $3894$
& $1.13 \times 10^{-8}$
& $1.05 \times 10^{-6}$
\\ 
$512$
& $864$
& $192$
& $0.5$
& $280$M
& $0.4$M
& $2.4$k
& $5226$
& $4.05 \times 10^{-9}$
& $5.10 \times 10^{-7}$
\\\hline
\end{tabular*}
\end{table*}
\par
\begin{figure}[tb]
\centerline{\includegraphics[scale=0.5]{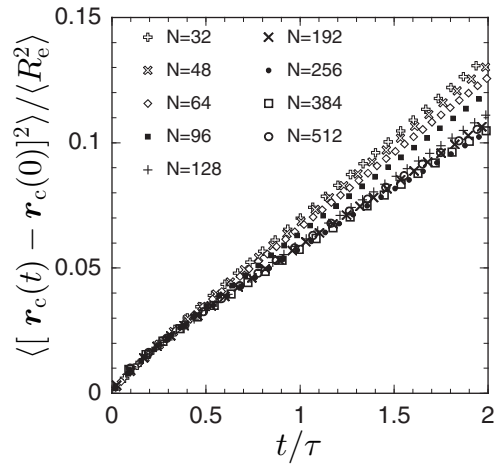}}
\caption{
The elapsed time $t$ dependence of
the mean square displacement
$\langle [ \mib{r}_{\rm c}(t) - \mib{r}_{\rm c}(0) ]^2 \rangle$
of the center of mass of a polymer chain.
The mean square displacement and the elapsed time are scaled by
the mean square end-to-end distance $\langle R_{\rm e}^2 \rangle$
and the longest relaxation time $\tau$, respectively.
}
\label{msd}
\end{figure}

Figure \ref{msd} shows
the elapsed time dependence of
the mean square displacement
$\langle [ \mib{r}_{\rm c}(t) - \mib{r}_{\rm c}(0) ]^2 \rangle$
of the center of mass $\mib{r}_{\rm c}$ of a polymer chain,
where $t$ denotes the elapsed time.
In Fig.\ \ref{msd},
the mean square displacement and the elapsed time are scaled by
the mean square end-to-end distance $\langle R_{\rm e}^2 \rangle$
and the longest relaxation time $\tau$, respectively.
The values of
$\langle R_{\rm e}^2 \rangle$ and $\tau$
are shown in Table \ref{t-values}.
For $N \le 256$,
the values
estimated in
Ref.\ \citen%
{HagitaTakano:melt}
are used,
while 
those estimated from longer simulations
than in Ref.\ \citen%
{HagitaTakano:melt}
are used for $N \ge 384$.
Here,
the longest relaxation time $\tau$ is estimated
by the same method as in Ref.\ \citen%
{HagitaTakano:melt},
which is based on the variational estimation of
the relaxation modes and rates%
\cite{TakanoMiyashita}
and has been applied to various polymer systems.%
\cite{%
HagitaTakano:melt,%
KosekiHiraoTakano,%
HiraoKosekiTakano,%
HagitaTakano:tube,%
HagitaKosekiTakano:slit,%
HagitaIshizukaTakano%
}
The self-diffusion constant $D$ is estimated
by fitting the data points
at times longer than the longest relaxation time $\tau$
to a straight line
$\langle [ \mib{r}_{\rm c}(t) - \mib{r}_{\rm c}(0) ]^2 \rangle
= 6Dt + {\rm constant}$.
The estimated values of $D$ are shown in Table
\ref{t-values}.
\par
\begin{figure}[tb]
\centerline{\includegraphics[scale=0.5]{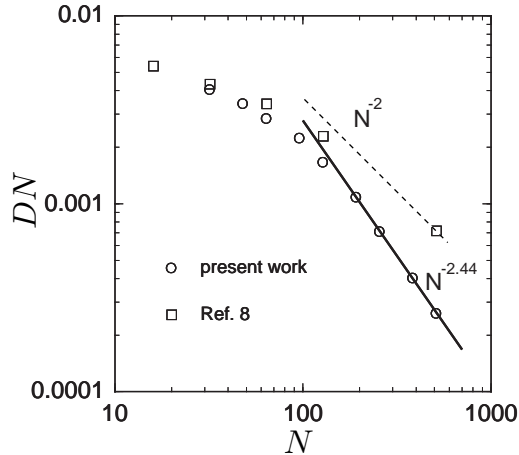}}
\caption{
Log-log plot of $DN$ versus $N$.
Solid line represents
the result of
the fit of the data for
$N= 256, 384$ and $512$ explained in the text.
The results of Ref.\ \citen{KreerBaschnagelMullerBinder}
are also shown for comparison.
Dashed line represents
the theoretical result $D \propto N^{-2}$.
}
\label{diff}
\end{figure}

Figure \ref{diff} shows
a log-log plot of $D N$ versus $N$.
The apparent exponent $x_{\rm d}$ of the power law
dependence of the self-diffusion constant
$D \propto N^{-x_{\rm d}}$,
which appears as the slope in the log-log plot,
increases as $N$ increases.
It is smaller than two for $N \le 96$
and larger than two for $N \ge 128$.
The least squares fit of the data points
for $N =128, 192$ and $256$
to a straight line in
a log-log plot of $D$ versus $N$
gives $x_{\rm d} \simeq 2.21$,
while that for $N =256, 384$ and $512$ gives $x_{\rm d} \simeq 2.44$.
The latter result is shown in Fig.\ \ref{diff} as a solid line.
\par
The reptation theory predicts that
$\tau \propto N^3$ and  $D \propto N^{-2}$
in the large $N$ limit,
while
$\langle R_{\rm e}^2 \rangle \propto N$ is expected to hold in a melt.
Thus,
the ratio
$ D \tau / \langle R_{\rm e}^2 \rangle $ 
is predicted to be
independent of $N$
for sufficiently large $N$.
From the values in Table \ref{t-values},
the apparent exponents 
of the power law dependences
$\langle R_{\rm e}^2 \rangle \propto N^{2\nu}$
and
$\tau \propto N^{x_{\rm r}}$
are estimated
as $2\nu \simeq 1.02$ and $x_{\rm r} \simeq 3.52$
for $N=256$, $384$ and $512$.
As mentioned before,
the apparent exponent $x_{\rm d}$
in
$D \propto N^{-x_{\rm d}}$
is estimated as
$x_{\rm d} \simeq 2.44$
for $N=256$, $384$ and $512$.
Thus,
the relation
$-x_{\rm d}+x_{\rm r}-2\nu \simeq 0$ seems to hold.
This means that
$ D \tau / \langle R_{\rm e}^2 \rangle $ 
is independent of $N$
even in the range of $N$
where
the true asymptotic behaviors of $\tau$ and $D$
are not expected to be seen.
This behavior can be seen
from the fact that
the slopes in Fig.\ \ref{msd},
which correspond to
$ 6 D \tau / \langle R_{\rm e}^2 \rangle $,
are almost the same for $N \ge 192$.
\par
\begin{figure}[tb]
\centerline{\includegraphics[scale=0.5]{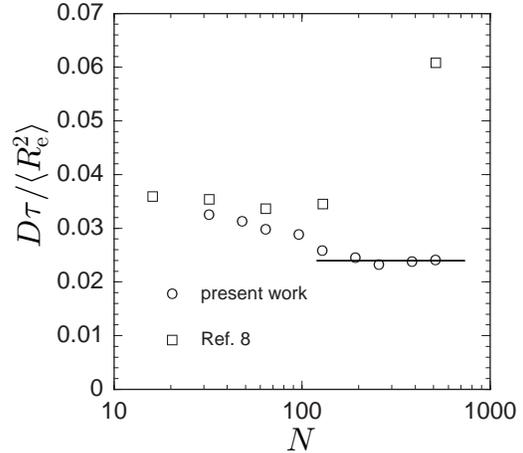}}
\caption{
Semi-log plot of
$D \tau / \langle R_{\rm e}^{2} \rangle$
versus $N$.
Solid line represents
the average of the values of
$D \tau / \langle R_{\rm e}^{2} \rangle$
for $N=192, 256, 384$ and $512$.
The results of Ref.\ 
\citen{KreerBaschnagelMullerBinder}
are also shown for comparison.
In the results of Ref.\ 
\citen{KreerBaschnagelMullerBinder},
$\tau_{p=1}$\cite{KreerBaschnagelMullerBinder}
is used as the longest relaxation time.
}
\label{fig3}
\end{figure}

According to the reptation theory,
the ratio
$ D \tau / \langle R_{\rm e}^2 \rangle $
is not only independent of $N$
but also
a constant which contains no adjustable parameter.
Its value is given by $ 1/(3 \pi^{2}) \simeq 0.034$.%
\cite{DoiEdwards}
Figure  \ref{fig3} shows a semi-log plot 
of $D \tau / \langle R_{\rm e}^{2} \rangle $
versus $N$.
For $N \geq 192$,
the values of
$D \tau / \langle R_{\rm e}^{2} \rangle$
seem to converge to a constant value around $0.024$.
If this value can be regarded as
the large $N$ limit of
$D \tau / \langle R_{\rm e}^{2} \rangle $,
the value is about thirty percent smaller than the value predicted by
the reptation theory.
The agreement between the two values
is considered to be reasonable
if we take into account the following facts.
Firstly, 
the estimated value of $\tau$
is considered to have
a tendency to be smaller than the exact value,
because
the relaxation time $\tau$ is estimated by the variational method.%
\cite{HagitaTakano:melt,TakanoMiyashita}
Secondly,
the accuracy of the predicted value of 
$D \tau / \langle R_{\rm e}^{2} \rangle$
itself is unknown,
since the reptation theory is based on
a simple model of polymer melts.
\par
In summary, 
we have examined the $N$-dependence of 
the self-diffusion constant $D$
of a polymer chain of $N$ segments in a melt
through Monte Carlo simulations of the bond fluctuation model.
The apparent exponent $x_{\rm d}$ in the power law
dependence $D \propto N^{-x_{\rm d}}$
is found to increase beyond two as $N$ increases.
From the data for $N=256$, $384$ and $512$,
$x_{\rm d}$
is estimated as $x_{\rm d} \simeq 2.44$.
For large values of $N$,
the ratio
$D \tau / \langle R_{\rm e}^{2} \rangle$
seems to be a constant value around 0.024,
which reasonably agrees with the theoretical prediction.
The reptation theory predicts
that
$D \tau / \langle R_{\rm e}^{2} \rangle$
is independent of the monomer concentration of a melt.%
\cite{DoiEdwards}
Therefore, it is interesting to investigate the volume fraction $\phi$
dependences of
$D$, $\tau$ and $\langle R_{\rm e}^{2} \rangle$
in addition to their $N$-dependences.
A study in this direction is in progress
\par
The authors are grateful to
Professor S.\ F.\  Edwards,
Professor E.\ M.\ Terentjev,
Professor M.\ Doi and
Professor P.\ G.\ de Gennes
for their comments.
The present work was partially supported by 
a Grant-in-Aid for Scientific Research (C) and
a Grant-in-Aid for Scientific Research on Priority Areas
from the Ministry of Education, Culture, Sports, Science and Technology.
The authors thank
Research Center for Computational Science of
Okazaki National Research Institutes
for the use of the Fujitsu VPP 5000
and Hokkaido University Computing Center
for the use of the Hewlett-Packard V2500.
One of the authors (K.H.) thanks
Professor E.\ Hanamura
at 
Hanamura Team, 
CREST, Japan Science and Technology Corporation
and Chitose Institute of Science and Technology,
where a part of the present work was carried out.
\par\noindent

\end{document}